# Sequence Analysis of Learning Behavior in Different Consecutive Activities

Abdulelah Abuabat
School of Computing and Information Sciences
University of Pittsburgh
Pittsburgh, PA 15260, USA
aaa185@pitt.edu

Peter Brusilovsky
School of Computing and Information Sciences
University of Pittsburgh
Pittsburgh, PA 15260, USA
peterb@pitt.edu

**Abstract:** The purpose of this research is to study the possibility of identifying students, statistically, by analyzing their behavior in different consecutive activities. In this project, there are three different sorts of activities: animated example, basic example, and parameterized exercises. We extracted the behavior of each student from the log activities of the Mastery Grids platform. Additionally, we investigate by using unsupervised learning technique, whether there are common patterns, that students share or not while performing these activities. We conclude that we are able to identify students from their behavior, besides that there are some common patterns.

## Introduction

Day after day, research shows how vital sequential analysis is, especially since many settings depend on it. One of these settings is studying user behavior (student, customer, etc.) and trying to discover if there are concrete results and indications that can be used to help the user to gain more benefits from the system. Our interest lies in analyzing both independent and dependent activities sequentially, and checking what knowledge can be acquired, and then how to apply this knowledge to help the user at the end.

In this study, we are interested in examining sequence independent activities in an educational system, called Mastery Grids: a parameterized Java exercise (Hsiao IH, 2009). This system is a self-assessment education system for an introduction to the object-oriented programming course in the School of Computing and Information at the University of Pittsburgh, and the aim of this system is to help students to practice the theory that they have learned in the class. Our goal here is to check if users, in this system, can be identified by their behavior, and subsequently, how stable this behavior is overall. Additionally, we applied a hierarchal analysis technique to see if there are common behavior traits among students, and what behavior traits looks like. We were inspired by the work of (Guerra, 2014), where they study the behavior of the students in one type of activity, which is in this case parameterized exercise. However, to have more generalize and accurate conclusions, it is superior, in our view, to examine not just one type of activity, but different forms of consecutive activities to see whether the behaviors are stable or sporadic.

## Dataset

We used the log activities of the Mastery Grids. The logs stored the students' activities from opening the session until the student quit. In this analysis, we included all the students for two semesters, who had the following three activities: animated example, basic example, and parameterized exercises. So, we ended with a total of 44 students. The features that we considered from this log were: student username, session, topic, and the activity itself. The highest number of sessions in this log was 42, and the number of topics were 21.

**Activities Labeling**

In each sequence, we considered the duration of each activity, and the correctness if it is a parameterized exercise, in (Fig. 1). An example of a sequence is: "AnEx ex f P p", which means that the student spent more than the median time doing the animated example activity "AnEx". Then, the student spent less than the median time reviewing the basic example "ex" ... etc.,[1]. Obviously, we used the median since the values were not normally distributed. The sequence started and ended alongside the system session. Thus, if a student shifted between topics, while they were in the same system session, for each topic the sequence construction will not be interrupted. At the end, we recorded around 650 sequences.

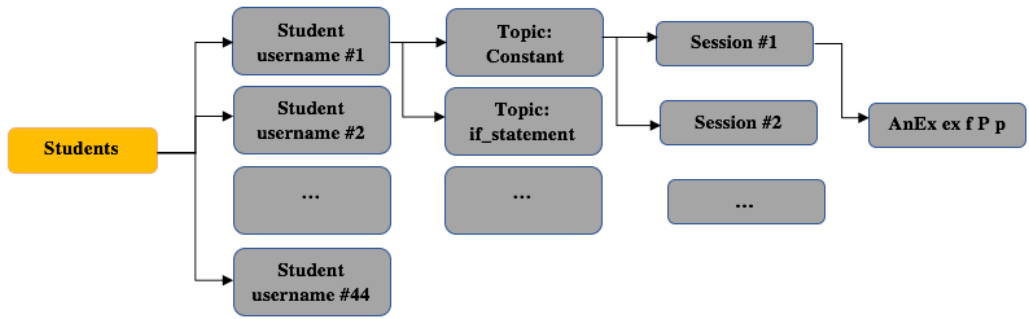

**Figure 1.** Labeling Structure.

Moreover, we tested different approaches in defining the start and end point for the sequence, and we will mention three examples. First, the time between activities must be less than the median of all the activities in order to make sure that the student was working without interruption. Second, each sequence should contain exercises besides the examples, not just one type of activity. Third, each sequence should end with exercises in order to see how valuable and influential the basic examples and the animated examples were for the exercise results. However, we did not get worthwhile results. The reason for that we believe is that these sorts of definitions reduce the amount of sequences pointedly, which affect the pattern mining process.

## Method

In this phase we examined the results of labeling those 650 sequences by conducting a sequential pattern mining. From the sequential pattern mining results, we studied two things: 1) The ability of recognizing the students based on their behavior or pattern while practicing different activities--study this, we checked the pattern stability for each student and 2) Clustering the students into two clusters and checking which patterns are common in each cluster.

**Sequential Pattern Mining Algorithm**

In order to discover the most frequent sequence and pattern among my sequences, we used the SPAM algorithm (Ayres, J, 2002), which is a sequential pattern mining using a bitmap representation[2]. In this algorithm, we have to decide two things: the minimum support "minsup" and the maximum gap. Minimum support is the support of a pattern $\chi$ which is the percentage of sequences of the total sequences in the dataset. This contains $\chi$ as a sequence or a sub- sequence. After several tests, and based on the results that we got, we found 4% was a practical choice for the minimum support, and 1 for the maximum gap. Lastly, we only considered patterns with 2 as the minimum length in order to better understand the patterns' results (Fig. 2).

---

[1] Labeling abbreviations:
ex = spend more than the median in the basic example.
Ex = spend less than the median in the basic example.
AnEx = spend more than the median in the animated example.
anex = spend less than the median in the animated example.
P = quiz - Pass - user spends more than the median.
F = quiz - Fail - user spends more than the median.
p = quiz - Pass - user spends less than the median.
f = quiz - Fail - user spends less than the median.
[2] http://www.philippe-fournier-viger.com/spmf/SPAM.php

**Pattern Stability**

To have enhanced insight about the results, we checked how many times each pattern occurred for each student. Then, we normalized the numbers. Thus, if a certain pattern did not occur, we smoothed it by storing it as a very small number, 0.0001, so we can get more sense when analyzing the results after calculating the distance in the next step.

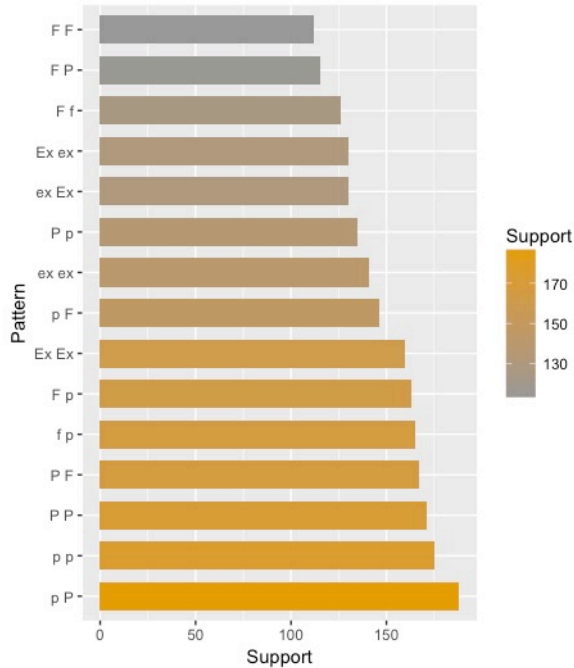

**Figure 2.** The Common 15 Patterns.

Regarding the stability test (Guerra, 2014), we split the sequence activities for each student randomly, and these two parts were represented as vectors. Next, we checked the distance between each student and their other part, and the distance between each student and other students' parts. If the pattern is stable, then the distance between each student's parts will be closer than the distance between each student with other students' parts. Since the situation here is symmetry, $sim(x, y) = sim(y, x)$, we have used both the symmetric version of Kullback-Leibler, which is Jensen-Shannon (JS) divergence (Majtey, 2005), and the cosine similarity measurement to calculate the distance between the two distributions. The results showed that the pattern was stable, and the pattern did not happen arbitrarily (Tab. 1), and the *self-distance* is considerably less than the *distance-to-other* in both measurements. Lastly, to check how significant the difference between the mean of the two populations: *self-distance* and *distance-to-other*, we applied a paired samples t-test, and the results in (Tab. 1) confirmed that the differences were not at random.

| Measure | Self-Distance | Distance-to-Other | t-test | p-value |
|---|---|---|---|---|
| Jensen-Shannon (JS) Divergence | 0.370 | 0.514 | -7.84 | < 0.001 |
| Cosine Similarity Measurement | 0.362 | 0.53 | -7.71 | < 0.001 |

**Table 1.** The Results of the Distance Between the Two Distributions

**Clustering Students Based on Their Pattern**

Another way to look at the patterns is by dividing the students into different clusters. This might help us to infer if there is a successful pattern or not, and if we can distinguish between a pattern that can lead to positive results and a pattern which can lead to undesirable results. We applied an unsupervised machine learning technique to give us an overview of the patterns, and to show us the general behavior for each group. In this context, we applied a Hierarchical Clustering, using the *Ward* method, with k = 2. (Fig. 3). And as you can see (Fig. 4), it seems that Cluster #1 spent more time practicing parameterized exercises. Furthermore, Cluster #1 tended to repeat the exercises and gain positive results. In contrast, Cluster #2 looks like they paid more attention to the basic examples and the animated examples.

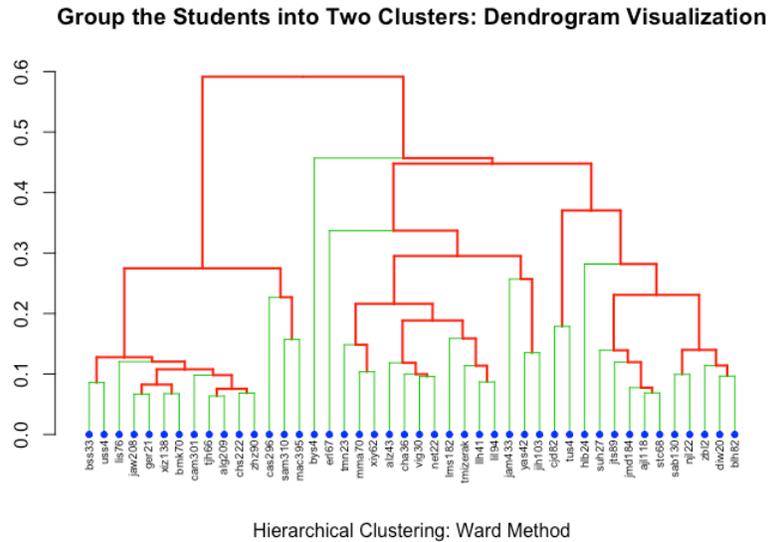

**Figure 3.** Dendrogram Visualization for the two Clusters by using Ward Method.

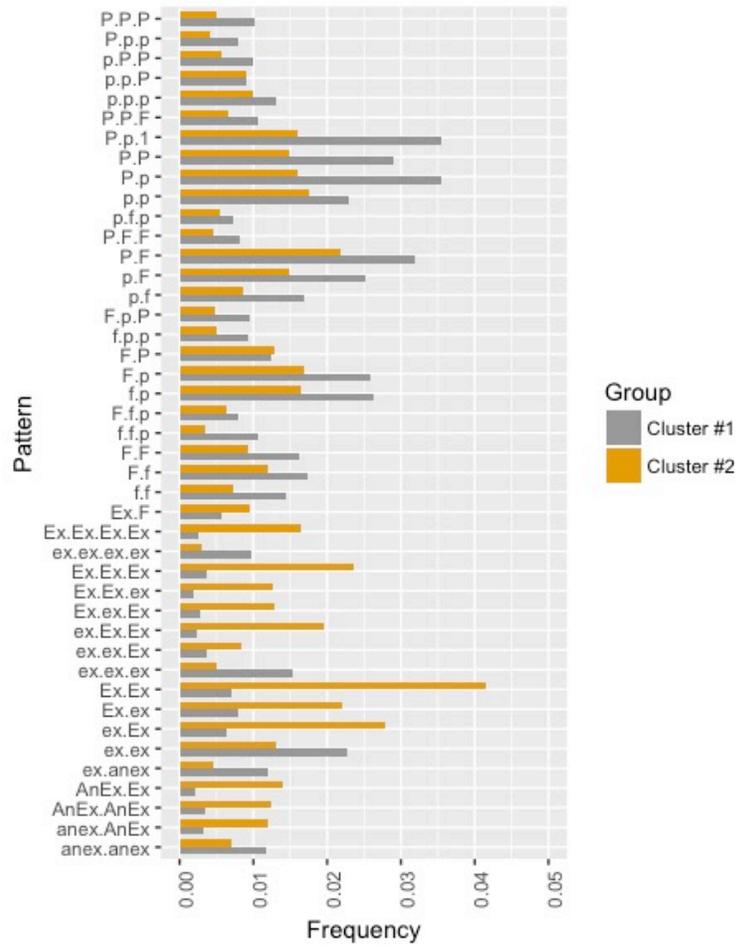

**Figure 4.** A Bar Chart that Shows the Frequency for each Pattern in each Cluster.

## Discussion and Conclusion

In this research, we investigated if we can identify students by their behavior in different consecutive activities. We found that students can be statistically identified, as we presented. Additionally, some students are sharing the same behavior, which could lead to studies of the final outcome of each pattern, for instance, or discover the most effective pattern.

The overall results of our research are promising, and it is possible to get more insightful and tangible knowledge regarding user behavior in such a system. Further research can also lead to the level of a heuristic system, which can guide the students in a more successful and effective manner. However, the dataset needs additional observations and extra features such as the final grade for the course in order to reach this level of intelligence. At the end, this study leads to additional research questions such as the extent that the user can be identified by his/her behavior, especially when performing a different type of activity, and if it is possible to change user behavior and steer it in a different direction.

# Acknowledgment

We would like to thank Jordan Barría for his great assistance and guidance, and for his help to review our results.